\documentclass[11pt,a4paper,twoside]{article}

\usepackage[T2A]{fontenc}
\usepackage[cp1251]{inputenc}
\usepackage[english]{babel}

\usepackage[psamsfonts]{amsfonts}
\usepackage{amsthm}
\usepackage{amssymb}
\usepackage{amsmath}
\usepackage{amsopn}
\usepackage{bm}
\usepackage{cite}

\newcommand{\nameABLemma}{$\mathcal{A}$-$\mathcal{B}$ bracket lemma}
\newcommand{\nameConjVLemma}{Conjugate variable lemma}
\newcommand{\nameSoV}{Separation of variables theorem}

\newtheorem{ABLemma}{\nameABLemma}{\bf}{\it}

\newtheorem{ConjVLemma}{\nameConjVLemma}{\bf}{\it}
\newtheorem{SoV}{\nameSoV}{\bf}{\it}

\theoremstyle{plain}
\newtheorem{proposition}{Proposition}[section]

\newtheorem*{proposition*}{Proposition}

\newtheorem{example}{Example} 
\newtheorem{remark}{Remark}

\newcommand{\tens}[1]{\textsf{#1}}
\renewcommand{\vec}[1]{\bm{#1}}

\DeclareMathOperator{\Complex}{\mathbb{C}}

\DeclareMathOperator{\Integer}{\mathbb{Z}}
\DeclareMathOperator{\Tr}{Tr} \DeclareMathOperator*{\res}{res}
 \DeclareMathOperator{\ad}{ad}
\DeclareMathOperator{\Sym}{Sym} 
\DeclareMathOperator{\const}{const}
\DeclareMathOperator{\diag}{diag}
\DeclareMathOperator{\Ibb}{\mathbb{I}}
\DeclareMathOperator{\rank}{rank}

\allowdisplaybreaks[4]

\begin{document}
\date{}
\title{\bf Orbit Approach to Separation of Variables \\ in $\mathfrak{sl}(4)$-Related Integrable Systems}

\author{\textbf{Julia Bernatska} \\ \small
 National University of 'Kyiv Mohyla Academy'\\ \small
 BernatskaJM@ukma.kiev.ua }

\maketitle
\begin{abstract}
Separation of variables by means of the orbit method is
implemented to integrable systems on coadjoint orbits in an $\mathfrak{sl}(4)$ loop algebra.
This is a development and a kind of explanation for Sklyanin's procedure of
separation of variables.
It is shown that points on a spectral curve serve as variables of separation
for two integrable systems living on two generic orbits embedded into a common manifold.
These orbits are endowed with different nonsingular Lie-Poisson brackets.
Explicit expressions for the case of $\mathfrak{sl}(4)$ loop algebra are given.
\end{abstract}

\section{Introduction}
Here we continue to develop the orbit approach to separation of variables
presented in \cite{BernHol2013}, where $\mathfrak{sl}(3)$-related integrable systems
are considered. Separation of variables in dynamical systems on coadjoint orbits
in an $\mathfrak{sl}(3)$ loop algebra are discussed in a number of papers, cited in \cite{BernHol2013}.
Now we consider the case of $\mathfrak{sl}(4)$ loop algebra, uninvestigated yet because of its
computational complexity. We try to overcome this complexity by means of the orbit approach.

\section{Preliminaries}
First we construct a loop algebra based
on the algebra $\mathfrak{g}\,{=}\,\mathfrak{sl}(4,\Complex)$ with the
Cartan-Weyl basis
\begin{gather*}
  \tens{X}_1=\tens{E}_{12}, \quad
  \tens{X}_2=\tens{E}_{23}, \quad
  \tens{X}_3=\tens{E}_{13}, \quad
  \tens{X}_4=\tens{E}_{34}, \quad
  \tens{X}_5=\tens{E}_{24}, \quad
  \tens{X}_6=\tens{E}_{14}, \\
  \tens{Y}_1=\tens{E}_{21}, \quad
  \tens{Y}_2=\tens{E}_{32}, \quad
  \tens{Y}_3=\tens{E}_{31}, \quad
  \tens{Y}_4=\tens{E}_{43}, \quad
  \tens{Y}_5=\tens{E}_{42}, \quad
  \tens{Y}_6=\tens{E}_{41}, \\
  \tens{H}_1=\tfrac{1}{4}\diag(3,-1,-1,-1),\quad
  \tens{H}_2=\tfrac{1}{2}\diag(1,1,-1,-1),\quad
  \tens{H}_3=\tfrac{1}{4}\diag(1,1,1,-3).
\end{gather*}
By $\tens{E}_{ij}$ we denote the standard basis in the vector space of $4\,{\times}\,4$ matrices, i.\;e.
$\tens{E}_{ij}$ is the matrix with only 1 at the position of row $i$ column $j$ and 0s at all other positions.
The matrices $\tens{H}_1$, $\tens{H}_2$, $\tens{H}_3$ are chosen to be the dual basis to the standard
basis in the Cartan subalgebra:
$$\tens{H}_1^{\ast}=\diag(1,-1,0,0),\quad \tens{H}_2^{\ast}=\diag(0,1,-1,0),\quad
\tens{H}_3^{\ast}=\diag(0,0,1,-1)$$
with respect to the bilinear form $\langle \tens{A}, \tens{B} \rangle\,{=}\,\Tr \tens{A} \tens{B}$.
In what follows we denote the set $\{\tens{H}_1$, $\tens{Y}_1$, $\tens{X}_1$, $\tens{Y}_3$, $\tens{X}_3$, $\tens{Y}_6$, $\tens{X}_6$, $\tens{H}_2$, $\tens{Y}_2$, $\tens{X}_2,\,\tens{Y}_5,\,\tens{X}_5,\,
\tens{H}_3,\,\tens{Y}_4,\,\tens{X}_4\}$  by $\{\tens{Z}_a\}_{a=1}^{15}$.
We also introduce the dual algebra
$\mathfrak{g}^{\ast}$ with respect to the mentioned bilinear form, we denote its basis by $\{\tens{Z}_a^{\ast}\}$:
$$\tens{X}_j^{\ast}=\tens{Y}_j,\ \ \tens{Y}_j^{\ast}=\tens{X}_j,\ j=1,\,\dots,\,6,\quad \tens{H}_1^{\ast},\quad
\tens{H}_2^{\ast},\quad \tens{H}_3^{\ast}.$$

Let $\mathcal{P}(\lambda,\lambda^{-1})$ be the algebra of
Laurent polynomials in $\lambda$, and
$\widetilde{\mathfrak{g}}$ be the loop algebra
$\mathfrak{sl}(4,\Complex)\otimes
\mathcal{P}(\lambda,\lambda^{-1})$. Then
\begin{equation*}
  \tens{Z}_a^{m} = \lambda^m \tens{Z}_a,\qquad m\in\Integer,\quad a=1,\,\dots,\,15
\end{equation*}
form a basis in $\widetilde{\mathfrak{g}}$. This loop algebra
is standard graded with respect to
the operator $\mathfrak{d} \,{=}\,d/d\lambda$ of homogeneous degree.
By $\mathfrak{g}_m$, $m\,{\in}\, \Integer$ we denote the
eigenspace of degree~$m$.

According to the Kostant-Adler scheme \cite{AdlerMoerb}
$\widetilde{\mathfrak{g}}$ is decomposed into two subalgebras
\begin{equation*}
  \widetilde{\mathfrak{g}}_+ = \sum_{m\geqslant
  0}\mathfrak{g}_m,\qquad
  \widetilde{\mathfrak{g}}_- = \sum_{m < 0}\mathfrak{g}_m,\qquad
  \widetilde{\mathfrak{g}} = \widetilde{\mathfrak{g}}_+ +
  \widetilde{\mathfrak{g}}_-.
\end{equation*}
Further, we introduce the \emph{$\ad$-invariant bilinear forms}
\begin{equation*}
  \langle \tens{A}(\lambda), \tens{B}(\lambda) \rangle_{k} =
  \res_{\lambda=0} \lambda^{-k-1} \Tr \tens{A}(\lambda)
  \tens{B}(\lambda), \quad \tens{A}(\lambda),\ \tens{B}(\lambda)\in
  \widetilde{\mathfrak{g}}, \quad k\in \Integer
\end{equation*}
and use them to define the spaces dual to
$\widetilde{\mathfrak{g}}_+$ and $\widetilde{\mathfrak{g}}_-$.
\begin{example}\label{E:k-1}
With respect to the bilinear form $\langle \tens{A}(\lambda), \tens{B}(\lambda) \rangle_{-1} \,{=}\,
  \res_{\lambda=0} \Tr \tens{A}(\lambda) \tens{B}(\lambda)$  we obtain the following dual spaces
\begin{equation*}
  (\widetilde{\mathfrak{g}}_-)^{\ast} = \widetilde{\mathfrak{g}}_+ ,\qquad
  (\widetilde{\mathfrak{g}}_+)^{\ast} = \widetilde{\mathfrak{g}}_-,
\end{equation*}
where $(\widetilde{\mathfrak{g}}_-)^{\ast}$ and
$(\widetilde{\mathfrak{g}}_+)^{\ast}$ contain only nonzero
functionals on $\widetilde{\mathfrak{g}}_{\pm}$.
\end{example}
\begin{example}\label{E:kN}
With respect to the bilinear form $\langle \tens{A}(\lambda), \tens{B}(\lambda) \rangle_{N-1} \,{=}\,
\lambda^{-N} \res_{\lambda=0} \Tr \tens{A}(\lambda) \tens{B}(\lambda)$  we obtain the dual spaces
\begin{equation*}
  (\widetilde{\mathfrak{g}}_-)^{\ast} = \sum_{m\geqslant  N}
  \mathfrak{g}_{m},\qquad  (\widetilde{\mathfrak{g}}_+)^{\ast}=
  \sum_{m <  N} \mathfrak{g}_m.
\end{equation*}
\end{example}
Further we consider a subset of $\widetilde{\mathfrak{g}}_+$ as a dual space to the both of subalgebras:
$\widetilde{\mathfrak{g}}_+$ and $\widetilde{\mathfrak{g}}_-$. We are interested in its foliations into  orbits
of the coadjoint action of these two subalgebras.

\section{Orbits of $\mathfrak{sl}(4,\Complex)\otimes \mathcal{P}(\lambda,\lambda^{-1})$ as phase spaces}
Fixing $ N\,{\geqslant}\, 0$ we introduce the variables $\{\alpha_1^{(m)}$, $\beta_1^{(m)}$, $\gamma_1^{(m)}$, $\beta_3^{(m)}$, $\gamma_3^{(m)}$, $\beta_6^{(m)}$, $\gamma_6^{(m)}$,
$\alpha_2^{(m)}$, $\beta_2^{(m)}$, $\gamma_2^{(m)}$, $\beta_5^{(m)}$, $\gamma_5^{(m)}$,
$\alpha_3^{(m)}$, $\beta_4^{(m)}$, $\gamma_4^{(m)}\,{:}$ $m\,{=}\,0,\,1,\,\dots,\, N\}$
denoted all together by $\{L_a^{(m)}\}_{a=1}^{15}$. Consider a space
$\mathcal{M}\,{\in}\,\widetilde{\mathfrak{g}}^{\ast}$ of the elements
\begin{gather}\label{muExpr}
  \tens{L}(\lambda) = \small
  \sum_{m=0}^{ N} \sum_{a=1}^{\rank \mathfrak{g}} L_a^{(m)} \big(\tens{Z}_a^m\big)^\ast=  \begin{pmatrix}
  \alpha_1(\lambda) & \beta_1(\lambda) & \beta_3(\lambda) & \beta_6(\lambda) \\ \gamma_1(\lambda) &
  \alpha_2(\lambda)-\alpha_1(\lambda) & \beta_2(\lambda) & \beta_5(\lambda) \\ \gamma_3(\lambda) &
  \gamma_2(\lambda) & \alpha_3(\lambda)-\alpha_2(\lambda) & \beta_4(\lambda) \\
  \gamma_6(\lambda) & \gamma_5(\lambda) & \gamma_4(\lambda) & -\alpha_3(\lambda) \end{pmatrix},
\intertext{where} \nonumber
L_a(\lambda)=\sum_{m=0}^ N \lambda^m L_{a}^{(m)}.
\end{gather}
Let $\mathcal{C}(\mathcal{M})$ be the space of smooth functions on $\mathcal{M}$. For
all $f_1$, $f_2\,{\in}\, \mathcal{C}(\mathcal{M})$ we define the \emph{first Lie-Poisson
bracket} by the formula
\begin{gather}\label{LiePoissonBraNLS}
  \{f_1,f_2\}_{\text{f}} = \sum_{m,n=0}^ N \sum_{a,b=1}^{\rank \mathfrak{g}} P_{ab}^{mn}(-1)
  \frac{\partial f_1}{\partial L_a^{(m)}}\frac{\partial f_2}{\partial L_b^{(n)}},\\
 P_{ab}^{mn}(-1) = \langle \tens{L}(\lambda),
  [\tens{Z}_a^{-m-1},\tens{Z}_b^{-n-1}] \rangle_{-1}, \nonumber
\end{gather}
and turn the space $\mathcal{C}(\mathcal{M})$ into a phase space denoted by $\mathcal{D}_{\text{f}}$.
The variables $\{L_a^{( N)}\}$ annihilate the bracket \eqref{LiePoissonBraNLS},
therefore we consider them as constant in $\mathcal{D}_{\text{f}}$.
To make the introduced bracket nonsingular we restrict it to the subspace
$\mathcal{M}_0$ of $\mathcal{M}$ defined by the constraints
$$L_a^{( N)} = \const,\qquad a\,{=}\,1,\,\dots,\,15.$$
For all $f_1$, $f_2\,{\in}\, \mathcal{C}(\mathcal{M}_0)$ we define the \emph{second Lie-Poisson
bracket}  by the formula
\begin{gather}\label{LiePoissonBraHM}
  \{f_1,f_2\}_\text{s} = \sum_{m,n=0}^ N \sum_{a,b=1}^{\rank \mathfrak{g}} P_{ab}^{mn}( N-1)
  \frac{\partial f_1}{\partial L_a^{(m)}}\frac{\partial
  f_2}{\partial L_b^{(n)}},  \\ P_{ab}^{mn}( N-1) = \langle\tens{L}(\lambda),
   [\tens{Z}_a^{-m+ N-1},\tens{Z}_b^{-n+ N-1}] \rangle_{ N-1}\notag
\end{gather}
and introduce another phase space denoted by $\mathcal{D}_{\text{s}}$.
In what follows we consider the space of smooth functions $\mathcal{C}(\mathcal{M}_0)$, and use the set
$\{L_{a}^{(m)}\,{\mid}\,m\,{=}\,1,\,\dots,\,N\,{-}\,1\}$ as \emph{dynamic variables} in it. We call $\mathcal{M}_0$ a \emph{finite gap sector of $\widetilde{\mathfrak{g}}$}, more precisely
the $ N$-gap sector.

\begin{remark}
In addition to the brackets (\ref{LiePoissonBraNLS}) and
(\ref{LiePoissonBraHM}), one can define intermediate brackets
with the Poisson tensors
\begin{equation}\label{LiePoissonBras}
 P_{ab}^{mn}(k)=\langle \tens{L}(\lambda),
   [\tens{Z}_a^{-m+k},\tens{Z}_b^{-n+k}] \rangle_{k}, \qquad k=0, \ldots,  N-2.
\end{equation}
\end{remark}

According to Example~\ref{E:k-1} we consider $\mathcal{M}_0$
as located in $(\widetilde{\mathfrak{g}}_-)^{\ast}$ with respect to the bilinear form $\langle\cdot,\cdot \rangle_{-1}$.
Obviously, $\mathcal{M}_0$ is $\ad^\ast$-invariant under the coadjoint action of
the factor-algebra $\widetilde{\mathfrak{g}}_-/\sum_{l< {-} N} \mathfrak{g}_l$. In this connection we introduce
the first Lie-Poisson bracket $\{\cdot,\cdot\}_{\text{f}}$.

On the other hand,
according to Example~\ref{E:kN} we consider $\mathcal{M}_0$
as located in $(\widetilde{\mathfrak{g}}_+)^{\ast}$ with respect to the bilinear
form $\langle\cdot,\cdot \rangle_{ N-1}$. One can see that $\mathcal{M}_0$ is $\ad^\ast$-invariant
also under the coadjoint action of the factor-algebra
$\widetilde{\mathfrak{g}}_+/\sum_{l\geqslant  N} \mathfrak{g}_l$. So we introduce
the second Lie-Poisson bracket $\{\cdot,\cdot\}_{\text{s}}$.

Next, we introduce the following $\ad^{\ast}$-invariant functions in $\lambda$
(for the sake of simplicity we will often omit writing the dependence on $\lambda$)
\begin{align}
  &I_2(\lambda) \equiv \tfrac{1}{2} \Tr \tens{L}^2(\lambda) = \alpha_1^2 + \alpha_2^2 + \alpha_3^2 -
  \alpha_1\alpha_2 - \alpha_2\alpha_3 + \beta_1\gamma_1 + \beta_2\gamma_2 + \beta_3\gamma_3 + \beta_4\gamma_4
   + \beta_5\gamma_5 + \beta_6\gamma_6 = \notag \\
   &\phantom{I_2} = \small
  -\begin{vmatrix} \alpha_1 & \beta_1   \\
  \gamma_1 & \alpha_2-\alpha_1 \end{vmatrix} -
  \begin{vmatrix} \alpha_2-\alpha_1 & \beta_2  \\
  \gamma_2 & \alpha_3-\alpha_2 \end{vmatrix} -
  \begin{vmatrix} \alpha_1 & \beta_3  \\
  \gamma_3 & \alpha_3-\alpha_2 \end{vmatrix} -
  \begin{vmatrix} \alpha_3-\alpha_2 & \beta_4 \\ \gamma_4 & -\alpha_3 \end{vmatrix} - \notag \\
  &\phantom{I_2=} - \small
  \begin{vmatrix} \alpha_2-\alpha_1 & \beta_5 \\ \gamma_5 & -\alpha_3 \end{vmatrix}-
  \begin{vmatrix} \alpha_1 & \beta_6 \\ \gamma_6 & -\alpha_3 \end{vmatrix}, \label{InvarF} \\
&I_3(\lambda) \equiv\tfrac{1}{3} \Tr \tens{L}^3(\lambda) = \alpha_2\alpha_1^2 -
  \alpha_1 \alpha_2^2 + \alpha_3\alpha_2^2 - \alpha_2 \alpha_3^2 +
  \beta_1\gamma_1\alpha_2 + \beta_2\gamma_2(\alpha_3-\alpha_1) + \beta_3\gamma_3(\alpha_1-\alpha_2+\alpha_3)+ \notag
  \\ &\phantom{I_2} + \beta_3 \gamma_1 \gamma_2 + \beta_1 \beta_2 \gamma_3
  + \beta_4\big[\beta_2\gamma_5+\beta_3\gamma_6-\alpha_2\gamma_4\big] +
  \beta_5\big[\gamma_2\gamma_4+\beta_1\gamma_6-(\alpha_1-\alpha_2+\alpha_3)\gamma_5\big] + \notag
  \\ &\phantom{I_2}
  + \beta_6\big[\gamma_3\gamma_4+\gamma_1\gamma_5-(\alpha_3-\alpha_1)\gamma_6\big] = \small
  \begin{vmatrix} \alpha_1 & \beta_1 & \beta_3 \\
  \gamma_1 & \alpha_2-\alpha_1 & \beta_2 \\
  \gamma_3 & \gamma_2 & \alpha_3-\alpha_2 \end{vmatrix} +
  \begin{vmatrix} \alpha_1 & \beta_1 & \beta_6 \\
  \gamma_1 & \alpha_2-\alpha_1 & \beta_5 \\
  \gamma_6 & \gamma_5 & -\alpha_3 \end{vmatrix} + \notag
  \\ &\phantom{I_2} + \small \begin{vmatrix} \alpha_1 & \beta_3 & \beta_6 \\
  \gamma_3 & \alpha_3-\alpha_2 & \beta_4 \\  \gamma_6 & \gamma_4 & -\alpha_3 \end{vmatrix} +
  \begin{vmatrix} \alpha_2-\alpha_1 & \beta_2 & \beta_5 \\
  \gamma_2 & \alpha_3-\alpha_2 & \beta_4 \\  \gamma_5 & \gamma_4 & -\alpha_3 \end{vmatrix}, \notag \\
&I_4(\lambda)\equiv\tfrac{1}{4}\big[ \Tr \tens{L}^4(\lambda) -
\tfrac{1}{2}\big(\Tr \tens{L}^2(\lambda)\big)^2\big] = \small
 \alpha_3\begin{vmatrix} \alpha_1 & \beta_1 & \beta_3 \\
  \gamma_1 & \alpha_2-\alpha_1 & \beta_2 \\
  \gamma_3 & \gamma_2 & \alpha_3-\alpha_2 \end{vmatrix} +\beta_4
  \begin{vmatrix} \alpha_1 & \beta_1 & \beta_3 \\
  \gamma_1 & \alpha_2-\alpha_1 & \beta_2 \\
  \gamma_6 & \gamma_5 & \gamma_4 \end{vmatrix} - \notag \\ &\phantom{I_4} \small
  - \beta_5 \begin{vmatrix} \alpha_1 & \beta_1 & \beta_3 \\
  \gamma_3 & \gamma_2 & \alpha_3-\alpha_2 \\
  \gamma_6 & \gamma_5 & \gamma_4 \end{vmatrix} +\beta_6
  \begin{vmatrix} \gamma_1 & \alpha_2-\alpha_1 & \beta_2 \\
  \gamma_3 & \gamma_2 & \alpha_3-\alpha_2 \\
  \gamma_6 & \gamma_5 & \gamma_4 \end{vmatrix}
  = -\det \tens{L}(\lambda).
\end{align}
Every function $I_k$ is a sum of the diagonal minors of order $k$ up to a sign. The functions $\{I_1\,{=}\,0$, $I_2$, $\dots$, $I_{\rank \mathfrak{g}}\}$ serve as
coefficients of the characteristic polynomial of the $\tens{L}$-matrix, in our case:
$$\chi(w) = w^4- I_2 w^2 - I_3 w - I_4.$$
The functions $I_2$, $I_3$, $I_4$ are polynomials in the spectral parameter $\lambda$,
and their coefficients serve as invariant functions in dynamic variables, namely:
\begin{align}
  &I_2(\lambda)= h^{(0)}_2 + h^{(1)}_2 \lambda + \cdots + h^{(2N)}_{2}\lambda^{2N},\quad
  I_3(\lambda)= h^{(0)}_3 + h^{(1)}_3 \lambda + \cdots + h^{(3N)}_{3}\lambda^{3N},
  \\ & I_4(\lambda)= h^{(0)}_4 + h^{(1)}_4 \lambda + \cdots + h^{(4N)}_{4}\lambda^{4N}, \notag \\
  &h^{(\nu)}_{2} = \small -\sum_{m+n=\nu} \left( \begin{vmatrix} \alpha_1^{(m)} & \beta_1^{(n)}  \\
  \gamma_1^{(m)} & \alpha_2^{(n)}-\alpha_1^{(n)} \end{vmatrix} +
  \begin{vmatrix} \alpha_2^{(m)}-\alpha_1^{(m)} & \beta_2^{(n)} \notag \\
  \gamma_2^{(m)} & \alpha_3^{(n)}-\alpha_2^{(n)} \end{vmatrix} +
  \begin{vmatrix} \alpha_1^{(m)} & \beta_3^{(n)} \\
  \gamma_3^{(m)} & \alpha_3^{(n)}-\alpha_2^{(n)} \end{vmatrix} -\right. \\
  & \left. - \small\begin{vmatrix} \alpha_3^{(m)}-\alpha_2^{(m)} & \beta_4^{(n)}  \\
  \gamma_4^{(m)} & -\alpha_3^{(n)} \end{vmatrix} -
  \begin{vmatrix} \alpha^{(m)}_2-\alpha_1^{(m)} & \beta_5^{(n)}  \\
  \gamma_5^{(m)} & -\alpha_3^{(n)} \end{vmatrix} -
  \begin{vmatrix} \alpha_1^{(m)} & \beta_6^{(n)} \\
  \gamma_6^{(m)} & -\alpha_3^{(n)} \end{vmatrix} \right), \notag \\
  &\phantom{I_2(\lambda)} \nu=0,\,1,\,\dots,\,2 N; \label{InvFunc}  \\
  &h^{(\nu)}_{3} = \sum_{m+n+k=\nu} \left(\small
  \begin{vmatrix} \alpha_1^{(m)} & \beta_1^{(n)} & \beta_3^{(k)} \\
  \gamma_1^{(m)} & \alpha_2^{(n)}-\alpha_1^{(n)} & \beta_2^{(k)} \\
  \gamma_3^{(m)} & \gamma_2^{(n)} & \alpha_3^{(k)}-\alpha_2^{(k)} \end{vmatrix} +
  \begin{vmatrix} \alpha_1^{(m)} & \beta_1^{(n)} & \beta_6^{(k)} \\
  \gamma_1^{(m)} & \alpha_2^{(n)}-\alpha_1^{(n)} & \beta_5^{(k)} \\
  \gamma_6^{(m)} & \gamma_5^{(n)} & -\alpha_3^{(k)} \end{vmatrix} + \notag \right.
  \\ &\phantom{I_2} \left. + \small
  \begin{vmatrix} \alpha_1^{(m)} & \beta_3^{(n)} & \beta_6^{(k)} \\
  \gamma_3^{(m)} & \alpha_3^{(n)}-\alpha_2^{(n)} & \beta_4^{(k)} \\
  \gamma_6^{(m)} & \gamma_4^{(n)} & -\alpha_3^{(k)} \end{vmatrix} +
  \begin{vmatrix} \alpha_2^{(m)}-\alpha_1^{(m)} & \beta_2^{(n)} & \beta_5^{(k)} \\
  \gamma_2^{(m)} & \alpha_3^{(n)}-\alpha_2^{(n)} & \beta_4^{(k)} \\
  \gamma_5^{(m)} & \gamma_4^{(n)} & -\alpha_3^{(k)} \end{vmatrix} \right),
  \quad \nu=0,\,1,\,\dots,\,3 N, \notag \\
  &h^{(\nu)}_{4} = -\sum_{\substack{m+n+\\ +k+j=\nu}} \small
  \begin{vmatrix} \alpha_1^{(m)} & \beta_1^{(n)} & \beta_3^{(k)} & \beta_6^{(j)} \\
  \gamma_1^{(m)} & \alpha_2^{(n)}-\alpha_1^{(n)} & \beta_2^{(k)} & \beta_5^{(j)} \\
  \gamma_3^{(m)} & \gamma_2^{(n)} & \alpha_3^{(k)}-\alpha_2^{(k)} & \beta_4^{(j)} \\
  \gamma_6^{(m)} & \gamma_5^{(n)} & \beta_4^{(k)} & -\alpha_3^{(j)}
  \end{vmatrix},\quad \nu=0,\,1,\,\dots,\,4 N. \notag
\end{align}
Evidently, $h^{(2N)}_{2}$, $h^{(3N)}_{3}$, $h^{(4N)}_{4}$ are constant, for they do not contain dynamic variables.

The following assertions are immediately derived from the
Kostant-Adler scheme \cite{AdlerMoerb}.
\begin{proposition}
All functions $h^{(0)}_{2}$, $h^{(1)}_{2}$,\ldots, $h^{(2N-1)}_{2}$,
$h^{(0)}_{3}$, $h^{(1)}_{3}$,\ldots, $h^{(3N-1)}_{3}$,
$h^{(0)}_{4}$, $h^{(1)}_{4}$,\ldots, $h^{(4N-1)}_{4}$ defined by
\eqref{InvFunc} are in involution with respect to the brackets
\eqref{LiePoissonBraNLS} and \eqref{LiePoissonBraHM}.
\end{proposition}
\begin{proposition}\label{P:BraNLSannihil}
The functions $h^{(\nu)}_{2}$, $h^{(\nu+N)}_{3}$, $h^{(\nu+2N)}_{4}$, $\nu\,{=}\,N,\,\dots,\,  2N\,{-}\,1$
are functionally independent on
$\mathcal{M}_0$ and annihilate the first Lie-Poisson bracket \eqref{LiePoissonBraNLS}.
\end{proposition}
Let $\mathcal{O}_{\text{f}}\,{\subset}\,
\mathcal{M}_0$ be the algebraic manifold defined by
\begin{equation}\label{OrbEqNLS}
h^{(\nu)}_{2}\,{=}\,c^{(\nu)}_{2},\quad
h^{(\nu+ N)}_{3}\,{=}\,c^{(\nu+N)}_{3},\quad
h^{(\nu+2N)}_{4}\,{=}\,c^{(\nu+2N)}_{4},\qquad
\nu\,{=}\, N,\, \ldots,\, 2N\,{-}\,1,
\end{equation}
where all $c^{(\nu)}_{2}$, $c^{(\nu+N)}_{3}$, $c^{(\nu+2N)}_{4}$
are fixed complex numbers. The manifold $\mathcal{O}_{\text{f}}$
is a generic orbit of coadjoint action of the
subalgebra $\widetilde{\mathfrak{g}}_{-}$ on $\mathcal{M}_0$,
$\dim \mathcal{O}_{\text{f}}\,{=}\,12 N$.
Variation of the constants $c^{(\nu)}_{2}$, $c^{(\nu+N)}_{3}$, $c^{(\nu+2N)}_{4}$
gives a foliation of $\mathcal{M}_0$ into orbits of the first type.
Every orbit serves as a symplectic leaf in the symplectic manifold $\mathcal{M}_0$.

\begin{proposition}\label{P:BraHMannihil}
The functions $h^{(\nu)}_{2}$, $h^{(\nu)}_{3}$, $h^{(\nu)}_{4}$, $\nu\,{=}\,0,\,\dots,\,  N\,{-}\,1$ are functionally
independent on $\mathcal{M}_0$ and annihilate the second Lie-Poisson bracket~\eqref{LiePoissonBraHM}.
\end{proposition}
The algebraic manifold $\mathcal{O}_{\text{s}}\,{\subset}\,\mathcal{M}_0$ defined by
\begin{equation}\label{OrbEqHM}
h^{(\nu)}_{2}\,{=}\,c^{(\nu)}_{2},\quad
h^{(\nu)}_{3}\,{=}\,c^{(\nu)}_{3},\quad
h^{(\nu)}_{4}\,{=}\,c^{(\nu)}_{4},\qquad
\nu=0,\, \dots,\,  N\,{-}\,1,
\end{equation}
where all $c^{(\nu)}_{2}$, $c^{(\nu)}_{3}$, $c^{(\nu)}_{4}$ are fixed complex numbers,
is a generic orbit of coadjoint action of the
subalgebra $\widetilde{\mathfrak{g}}_{+}$ on $\mathcal{M}_0$,
$\dim \mathcal{O}_{\text{s}}\,{=}\,12 N$.
Variation of the constants $c^{(2)}_{\nu}$, $c^{(3)}_{\nu}$, $c^{(4)}_{\nu}$
gives another foliation of $\mathcal{M}_0$ into orbits of the second type.
In what follows we call $\mathcal{O}_{\text{f}}$ and $\mathcal{O}_{\text{s}}$ simply orbits,
and call \eqref{OrbEqNLS}, \eqref{OrbEqHM} orbit equations.

\section{Separation of variables on $\mathcal{O}_\text{f}$}
The orbit $\mathcal{O}_\text{f}$ with the first Lie-Poisson bracket \eqref{LiePoissonBraNLS}
has the following Poisson structure:
\begin{gather}\label{PoissonBraNLS}
\{L_{ij}^{(m)}, L_{kl}^{(n)}\}_\text{f} = L_{kj}^{(m+n+1)}\delta_{il} - L_{il}^{(m+n+1)} \delta_{kj},
\end{gather}
that can be written in terms of the $\tens{r}$-matrix
\begin{gather}
\{\tens{L}_1(u)\overset{\otimes}{,}\, \tens{L}_2(v)\}_\text{f} =
  [\tens{r}_{12}(u-v),\tens{L}_1(u)+\tens{L}_2(v)] \nonumber\\
\tens{r}_{12}(u-v) = \frac{1}{u-v}\, \sum_{a,b} \langle \tens{Z}_a,
\tens{Z}_b \rangle \tens{Z}_a^{\ast} \otimes \tens{Z}_b^{\ast}
\label{Rmatrix}
\end{gather}
with $\tens{L}_1(u)\,{=}\,\tens{L}(u)\,{\otimes}\, \Ibb$,
$\tens{L}_2(v)\,{=}\,\Ibb \,{\otimes}\, \tens{L}(v)$, where $\Ibb$ is the identity matrix.

In order to parameterize the orbit $\mathcal{O}_\text{f}$ of dimension $12N$ we need to eliminate $3N$
variables among $15N$ variables $\{L_a^{(m)}\}$. It is convenient to eliminate such set of variables that
all the invariant functions are linear in it. Such set of dynamic variables correspond to one of the maximal sets
of nilpotent commuting basis elements. Here we choose the variables $\{\beta_{4}^{(m)}, \beta_{5}^{(m)}, \beta_{6}^{(m)}\}$ for elimination.

Using linearity of the orbit equations \eqref{OrbEqNLS} in the chosen set of variables
one can write them in the matrix form
\begin{subequations}
\begin{gather} \label{ConstrNLS}
\vec{c} = \tens{F}^+ \vec{\beta} + \vec{a}^{+}, \\
\nonumber
\tens{F}^+ =\small \begin{bmatrix}
    \tens{F}_{ N} & \tens{F}_{ N-1} & \dots & \tens{F}_{1} & \tens{F}_{0} \\
   0 & \tens{F}_{ N} & \dots & \tens{F}_{2} & \tens{F}_{1} \\
   \vdots & \vdots& \ddots& \vdots& \vdots \\
   0 & 0 & \dots & \tens{F}_{ N} & \tens{F}_{ N-1} \\
   0 & 0 & \dots & 0 &  \tens{F}_{ N}
   \end{bmatrix},\quad
\vec{\beta} = \begin{bmatrix} \vec{\beta}^{(0)} \\ \vec{\beta}^{(1)} \\ \vdots \\
   \vec{\beta}^{(N-1)} \\ \vec{\beta}^{(N)}  \end{bmatrix}, \quad
\vec{c} = \begin{bmatrix} \vec{c}_{ N} \\ \vec{c}_{N+1} \\ \vdots \\
   \vec{c}_{2 N-1} \\ \vec{c}_{2 N} \end{bmatrix}, \quad
\vec{a}^{+} = \begin{bmatrix} \vec{a}_{N} \\
\vec{a}_{N+1} \\ \vdots \\
\vec{a}_{2N-1} \\ \vec{a}_{2 N} \end{bmatrix},\\ \small
\vec{\beta}^{(j)} = \begin{bmatrix} \beta_6^{(j)} \\ \beta_5^{(j)} \\ \beta_4^{(j)} \end{bmatrix}, \quad
\vec{c}_{j} = \begin{bmatrix} c^{(j)}_{2} \\ c^{(j+N)}_{3} \\ c^{(j+2N)}_{4} \end{bmatrix},\quad
\vec{a}_{j} = \begin{bmatrix} a^{(j)}_{2} \\ a^{(j+N)}_{3} \\ a^{(j+2N)}_{4} \end{bmatrix},\quad
\tens{F}_{j} = \begin{bmatrix}
B_{16}^{(j)} & B_{15}^{(j)} & B_{14}^{(j)} \\
B_{26}^{(j+N)} & B_{25}^{(j+N)} & B_{24}^{(j+N)} \\
B_{36}^{(j+2N)} & B_{35}^{(j+2N)} & B_{34}^{(j+2N)} \end{bmatrix}, \notag
\end{gather}
\begin{align}
&B_{16}^{(j)} =\gamma_6^{(j)},\quad B_{15}^{(j)} =\gamma_5^{(j)},\quad B_{14}^{(j)} =\gamma_4^{(j)},\notag \\
&B_{26}^{(j)} = \scriptsize \sum\limits_{\substack{m+\\+n=j}} \left(
   \begin{vmatrix} \gamma_1^{(m)} & \alpha_2^{(n)}\,{-}\,\alpha_1^{(n)} \\
  \gamma_6^{(m)} & \gamma_5^{(n)} \end{vmatrix} +
  \begin{vmatrix} \gamma_3^{(m)} & \alpha_3^{(n)}\,{-}\,\alpha_2^{(n)} \\
  \gamma_6^{(m)} & \gamma_4^{(n)} \end{vmatrix}\right), \
B_{36}^{(j)} = \scriptsize \sum\limits_{\substack{m+n+\\+k=j}}
  \begin{vmatrix} \gamma_1^{(m)} & \alpha_2^{(n)} \,{-}\, \alpha_1^{(n)} & \beta_2^{(k)} \\
  \gamma_3^{(m)} & \gamma_2^{(n)} & \alpha_3^{(k)} \,{-}\, \alpha_2^{(k)} \\
  \gamma_6^{(m)} & \gamma_5^{(n)} & \gamma_4^{(k)}  \end{vmatrix},\notag \\
&B_{25}^{(j)} = \scriptsize\sum\limits_{\substack{m+\\+n=j}} \left(-
   \begin{vmatrix} \alpha_1^{(m)} & \beta_1^{(n)}  \\
  \gamma_6^{(m)} & \gamma_5^{(n)} \end{vmatrix} +
  \begin{vmatrix} \gamma_2^{(m)} & \alpha_3^{(n)}-\alpha_2^{(n)}  \\
  \gamma_5^{(m)} & \gamma_4^{(n)} \end{vmatrix}\right), \quad
B_{35}^{(j)} = \scriptsize -\sum\limits_{\substack{m+n+\\+k=j}}
  \begin{vmatrix} \alpha_1^{(m)} & \beta_1^{(n)} & \beta_3^{(k)} \\
  \gamma_3^{(m)} & \gamma_2^{(n)} & \alpha_3^{(k)} - \alpha_2^{(k)} \\
  \gamma_6^{(m)} & \gamma_5^{(n)} & \gamma_4^{(k)}  \end{vmatrix},\label{BExpr}\\
&B_{24}^{(j)} = \scriptsize\sum\limits_{\substack{m+\\+n=j}} \left(-
  \begin{vmatrix} \alpha_1^{(m)} & \beta_3^{(n)} \\
  \gamma_6^{(m)} & \gamma_4^{(n)} \end{vmatrix} -
  \begin{vmatrix} \alpha_2^{(m)}-\alpha_1^{(m)} & \beta_2^{(n)}  \\
  \gamma_5^{(m)} & \gamma_4^{(n)} \end{vmatrix}\right), \quad
B_{34}^{(j)} = \scriptsize \sum\limits_{\substack{m+n+\\+k=j}}
  \begin{vmatrix} \alpha_1^{(m)} & \beta_1^{(n)} & \beta_3^{(k)} \\
  \gamma_1^{(m)} & \alpha_2^{(n)} - \alpha_1^{(n)} & \beta_2^{(k)} \\
  \gamma_6^{(m)} & \gamma_5^{(n)} & \gamma_4^{(k)}  \end{vmatrix},\notag \\
&A_{2}^{(j)} = \scriptsize\sum\limits_{\substack{m+\\+n=j}} \left(-
  \begin{vmatrix} \alpha_1^{(m)} & \beta_1^{(n)} \\
  \gamma_1^{(m)} & \alpha_2^{(n)}-\alpha_1^{(n)} \end{vmatrix} -
  \begin{vmatrix} \alpha_2^{(m)}-\alpha_1^{(m)} & \beta_2^{(n)}  \\
  \gamma_2^{(m)} & \alpha_3^{(n)}-\alpha_2^{(n)} \end{vmatrix} -
  \begin{vmatrix} \alpha_1^{(m)} & \beta_3^{(n)}  \\
  \gamma_3^{(m)} & \alpha_3^{(n)}-\alpha_2^{(n)} \end{vmatrix}\right), \notag \\
&A_{3}^{(j)} = \scriptsize \sum\limits_{\substack{m+n+\\+k=j}}
  \begin{vmatrix} \alpha_1^{(m)} & \beta_1^{(n)} & \beta_3^{(k)} \\
  \gamma_1^{(m)} & \alpha_2^{(n)} - \alpha_1^{(n)} & \beta_2^{(k)} \\
  \gamma_3^{(m)} & \gamma_2^{(n)} & \alpha_3^{(k)} - \alpha_2^{(k)} \end{vmatrix},\qquad
a^{(j)}_{2} = \scriptsize A_{2}^{(j)} + \sum\limits_{m+n=j} \alpha_3^{(m)}\alpha_3^{(n)}, \label{AExpr} \\
& a^{(j)}_{3} = A_{3}^{(j)} + \scriptsize \sum\limits_{m+n=j}  \alpha_3^{(m)} A_{2}^{(n)},\qquad
a^{(j)}_{4} = \scriptsize \sum\limits_{m+n=j} \alpha_3^{(m)} A_{3}^{(n)}. \notag
\end{align}
\end{subequations}
Supposing $\tens{F}_{N}$ is nonsingular, one easily eliminates the variables $\vec{\beta}$
\begin{gather*}
\vec{\beta} = (\tens{F}^+)^{-1} (\vec{c} - \vec{a}^{+}), \qquad
\text{or} \\ \small
\begin{bmatrix} \vec{\beta}_{0} \\ \vec{\beta}_{1} \\ \vdots \\
   \vec{\beta}_{ N-1} \\ \vec{\beta}_{ N}  \end{bmatrix} =
   \begin{bmatrix}
   \tens{F}_{ N}^{-1} & \widetilde{\tens{F}}_{ N-1} & \dots & \widetilde{\tens{F}}_{1} & \widetilde{\tens{F}}_{0}\\
   0 & \tens{F}_{ N}^{-1} & \dots & \widetilde{\tens{F}}_{2} & \widetilde{\tens{F}}_{1} \\
   \vdots & \vdots& \ddots& \vdots& \vdots \\
   0 & 0 & \dots & \tens{F}_{ N}^{-1} & \widetilde{\tens{F}}_{ N-1} \\
   0 & 0 & \dots & 0 & \tens{F}_{ N}^{-1}
   \end{bmatrix} \begin{bmatrix}
   \vec{c}_{ N} - \vec{a}_{ N} \\
   \vec{c}_{ N+1} - \vec{a}_{ N+1} \\ \vdots \\
   \vec{c}_{2 N-1} - \vec{a}_{2 N-1} \\
   \vec{c}_{2 N} - \vec{a}_{2 N} \end{bmatrix},\\
\widetilde{\tens{F}}_{ N-n} = \tens{F}_{ N}^{-1}
\sum_{k=1}^n \big(-\tens{F}_{ N-n-1+k} \tens{F}_{ N}^{-1}\big)^k, \qquad n=1,\,\dots,\, N.
\end{gather*}
Next, substitute $\vec{\beta}$ into  the Hamiltonians $h^{(0)}_{2}$, $h^{(1)}_{2}$,
\ldots, $h^{(N-1)}_{2}$, $h^{(0)}_{3}$, $h^{(1)}_{3}$,
\ldots, $h^{(2N-1)}_{3}$, $h^{(0)}_{4}$, $h^{(1)}_{4}$, \ldots, $h^{(3N-1)}_{4}$
\begin{equation}\label{HamiltNLS}
   \vec{h} = \tens{F}^{-} \vec{\beta} +  \vec{a}^{-} =
   \tens{F}^{-} (\tens{F}^{+})^{-1} \vec{c} + \vec{a}^{-} -
   \tens{F}^{-} (\tens{F}^{+})^{-1} \vec{a}^{+},
\end{equation}
where
\begin{gather*}
\tens{F}^{-} = \small \begin{bmatrix}
   \tens{g}_0 & 0 & \dots & 0 & 0 \\
   \tens{g}_1 & \tens{g}_0 & \dots & 0 & 0 \\
   \vdots & \vdots& \ddots& \vdots& \vdots \\
   \tens{g}_{ N-1} & \tens{g}_{ N-2} & \dots & \tens{g}_0 & 0 \\
   \tens{G}_{0} & \tens{g}_{ N-1}^0 & \dots & \tens{g}_1^0 & \tens{g}_0^0 \\
   \tens{G}_{1} & \tens{G}_{0} & \dots & \tens{g}_2^0 & \tens{g}_1^0 \\
   \vdots & \vdots& \ddots& \vdots& \vdots \\
   \tens{G}_{ N-1} & \tens{G}_{ N-2} & \dots &  \tens{g}_{0} & \tens{g}_{ N-1}^0\\
   \tens{F}_{0} & \tens{G}_{ N-1}^0 & \dots & \tens{G}_1^0 & \tens{G}_0^0 \\
   \tens{F}_{1} & \tens{F}_{0} & \dots & \tens{G}_2^0 & \tens{G}_1^0 \\
   \vdots & \vdots& \ddots& \vdots& \vdots \\
   \tens{F}_{ N-1} & \tens{F}_{ N-2} & \dots &  \tens{F}_{0} & \tens{G}_{ N-1}^0
   \end{bmatrix},\quad
\vec{h} = \begin{bmatrix} h^{(0)}_{4} \\ h^{(1)}_4 \\ \vdots \\  h^{(N-1)}_{4} \\
   \vec{\check{h}}_0 \\ \vec{\check{h}}_1 \\  \vdots \\ \vec{\check{h}}_{N-1}\\
   \vec{h}_0 \\ \vec{h}_1 \\  \vdots \\ \vec{h}_{N-1}
   \end{bmatrix},   \quad
\vec{a}^{-} = \begin{bmatrix} a^{(0)}_{4} \\ a^{(1)}_{4} \\ \vdots \\ a^{(N-1)}_{4} \\
\vec{\check{a}}_0 \\ \vec{\check{a}}_1 \\  \vdots \\ \vec{\check{a}}_{N-1} \\
\vec{a}_0 \\ \vec{a}_1 \\  \vdots \\ \vec{a}_{N-1}  \end{bmatrix},\\
\small
\tens{g}_j = \begin{bmatrix} B_{36}^{(j)} & B_{35}^{(j)} & B_{34}^{(j)} \end{bmatrix}, \quad
\tens{G}_j = \begin{bmatrix} B_{26}^{(j)} & B_{25}^{(j)} & B_{24}^{(j)} \\
B_{36}^{(j+N)} & B_{35}^{(j+N)} & B_{34}^{(j+N)} \end{bmatrix}, \\ \small
\tens{g}_j^0 = \begin{bmatrix} 0 & 0 & 0 \\ B_{36}^{(j)} & B_{35}^{(j)} & B_{34}^{(j)}\end{bmatrix}, \quad
\tens{G}_j^0 = \begin{bmatrix} 0 & 0 & 0 \\ B_{26}^{(j)} & B_{25}^{(j)} & B_{24}^{(j)} \\
B_{36}^{(j+N)} & B_{35}^{(j+N)} & B_{34}^{(j+N)} \end{bmatrix},\\ \small
\vec{\check{a}}_j = \begin{bmatrix} a^{(j)}_{3} \\ a^{(j+N)}_{4} \end{bmatrix},\quad
\vec{\check{h}}_j = \begin{bmatrix} h^{(j)}_{3} \\ h^{(j+N)}_{4} \end{bmatrix},\quad
\vec{h}_j = \begin{bmatrix} h^{(j)}_{2} \\ h^{(j+N)}_{3} \\ h^{(j+2N)}_{4} \end{bmatrix}.
\end{gather*}
Note that the expressions (\ref{HamiltNLS}) are linear in
$\{c^{(\nu)}_{2},\,c^{(\nu+N)}_{3},\,c^{(\nu+ 2N)}_{4}\,{\mid}\, \nu= N, \ldots, 2 N\}$.

To proceed we need to define the \emph{characteristic polynomial}
\begin{gather}\label{CharPoly}
  P(w,\lambda) = \det\bigl(\tens{L}(\lambda)- w \Ibb \bigr).
\end{gather}
It defines the spectral curve~$\mathcal{R}$:
\begin{equation}\label{SpectrCurve}
  w^4 - I_2(\lambda) w^2 - I_3(\lambda) w - I_4(\lambda) = 0,
\end{equation}
which is a curve of genus $4 N\,{-}3$ in general.
The spectral curve is common for
integrable systems over orbits of both types: $\mathcal{O}_\text{f}$ and $\mathcal{O}_\text{s}$.
Restriction to an orbit is realized through the orbit equations \eqref{OrbEqNLS}
or \eqref{OrbEqHM}, which fix some coefficients in \eqref{SpectrCurve}. The rest of coefficients
serve as Hamiltonians on an orbit and also remain constant during the evolution of a system.

Consider the spectral curve restricted to the orbit $\mathcal{O}_\text{f}$.
Denoting its points by $\{(\lambda_k,\,
w_k)\}$ we write the following set of equations for $6N$ Hamiltonians
\begin{multline}\label{HyperCurveNLS}
  w_k^4 = w_k^2\Big(h^{(0)}_{2}+h^{(1)}_2 \lambda_k + \cdots h^{(N-1)}_{2}\lambda_k^{N-1} +
  c^{(N)}_{2} \lambda_k^{N} + c^{(N+1)}_{2} \lambda_k^{N+1} + \cdots + c^{(2N)}_{2} \lambda_k^{2N}\Big)
  + \\ +w_k \Big(h^{(0)}_{3} + h^{(1)}_3 \lambda_k + \cdots h^{(2N-1)}_{3}\lambda_k^{2 N-1} +
  c^{(2N)}_{3} \lambda_k^{2N} + c^{(2N+1)}_{3} \lambda_k^{2N+1} + \cdots + c^{(3N)}_{3} \lambda_k^{3N}\Big) + \\
  + \Big(h^{(0)}_{4} + h^{(1)}_4 \lambda_k + \cdots h^{(3N-1)}_{4}\lambda_k^{3N-1} +
  c^{(3N)}_{4} \lambda_k^{3N} + c^{(3N+1)}_{4} \lambda_k^{3N+1} + \cdots + c^{(4N)}_{4} \lambda_k^{4N}\Big).
\end{multline}
So we need $6N$ points such that the system \eqref{HyperCurveNLS} is
uniquely solved for the Hamiltonians. Rewrite the above system of equations
in the matrix form
\begin{gather*}
\tens{W}^{-} \vec{h} +  \tens{W}^{+} \vec{c} = \vec{w}, \\
\small
\tens{W}^{-} =  \left[\begin{array}{cccccccccccc}
1 & \lambda_1 & \dots & \lambda_1^{ N-1} & \tens{W}_1^0 & \lambda_1 \tens{W}_1^0 & \dots & \lambda_1^{ N-1} \tens{W}_1^0
& \tens{W}_1 & \lambda_1 \tens{W}_1 & \dots & \lambda_1^{ N-1} \tens{W}_1 \\
1 & \lambda_2 & \dots & \lambda_2^{ N-1} & \tens{W}_2^0 & \lambda_2 \tens{W}_2^0 & \dots & \lambda_2^{ N-1} \tens{W}_2^0
& \tens{W}_2 & \lambda_2 \tens{W}_2 & \dots & \lambda_2^{ N-1} \tens{W}_2 \\
\vdots & \vdots & \dots & \vdots & \vdots & \vdots & \dots & \vdots \\
1 & \lambda_{6 N} & \dots & \lambda_{6 N}^{ N-1} &
\tens{W}_{6 N}^0 & \lambda_{6 N} \tens{W}_{6 N}^0 & \dots & \lambda_{6 N}^{ N-1} \tens{W}_{6 N}^0
& \tens{W}_{6 N} & \lambda_{6 N} \tens{W}_{6 N} & \dots & \lambda_{6 N}^{ N-1} \tens{W}_{6 N}\end{array}\right],\\
\tens{W}_k^0 \,{=}\, \begin{bmatrix} w_k & \lambda_k^{ N} \end{bmatrix},\quad
\tens{W}_k \,{=}\, \begin{bmatrix} w_k^2 & w_k\lambda_k^{ N} & \lambda_k^{2N} \end{bmatrix}\\
\tens{W}^{+} = \small \begin{bmatrix}
\lambda_1^{ N} \tens{W}_1 & \lambda_1^{ N+1} \tens{W}_1 & \dots & \lambda_1^{2 N} \tens{W}_1 \\
\lambda_2^{ N} \tens{W}_2 & \lambda_2^{ N+1} \tens{W}_2 & \dots & \lambda_2^{2 N} \tens{W}_2 \\
\vdots & \vdots & \dots & \vdots \\
\lambda_{6 N}^{ N} \tens{W}_{6 N} & \lambda_{6 N}^{ N+1} \tens{W}_{6 N}
& \dots & \lambda_{6 N}^{2 N} \tens{W}_{6 N}  \end{bmatrix},\qquad
\vec{w} = \begin{bmatrix} w_1^4 \\ w_2^4 \\ \vdots \\ w_{6 N}^4  \end{bmatrix}.
\end{gather*}
Suppose that all pairs $\{(\lambda_k,\, w_k)\,{\mid}$ $k\,{=}\,1$,
\ldots, $6 N\}$ are distinct points and $\tens{W}^-$ is nonsingular,
then the Hamiltonians can be expressed by the formula
\begin{equation}\label{hSystNLS}
 \vec{h} = -(\tens{W}^{-})^{-1} \tens{W}^{+} \vec{c} + (\tens{W}^{-})^{-1}\vec{w}.
\end{equation}
On the orbit $\mathcal{O}_\text{f}$ the formulas (\ref{HamiltNLS}) and (\ref{hSystNLS})
define the same set of functions, moreover, both
of them are linear in the parameters $\{c^{(\nu)}_{2},\,c^{(\nu+N)}_{3},\,c^{(\nu+2N)}_{4}\,{\mid}$
$\nu\,{=}\, N$, \ldots, $2 N\}$ of the orbit. These parameters are
independent, so one can equate the corresponding
terms, that is
\begin{gather}
 \tens{F}^{-} (\tens{F}^{+})^{-1} = - (\tens{W}^{-})^{-1} \tens{W}^{+},\qquad
 \vec{a}^{-} - \tens{F}^{-} (\tens{F}^{+})^{-1} \vec{a}^{+} =
 (\tens{W}^{-})^{-1}\vec{w}\quad \Rightarrow \nonumber\\
 \tens{W}^{-} \tens{F}^{-}  + \tens{W}^{+} \tens{F}^{+} = 0,\qquad
 \tens{W}^{-} \vec{a}^{-} +
 \tens{W}^{+} \vec{a}^{+} = \vec{w}. \label{RootEqNLS}
\end{gather}
The first matrix equation of \eqref{RootEqNLS} gives the following
\begin{gather}\label{Minors23}
\begin{bmatrix} B_{16}(\lambda_k) & B_{26}(\lambda_k) & B_{36}(\lambda_k)  \\
B_{15}(\lambda_k) & B_{25}(\lambda_k) & B_{35}(\lambda_k)\\
B_{14}(\lambda_k) & B_{24}(\lambda_k) & B_{34}(\lambda_k) \end{bmatrix}
\begin{bmatrix} w_k^2 \\ w_k \\ 1 \end{bmatrix} = \bm{0}.
\end{gather}
where the entries are polynomials in $\lambda_k$ with the coefficients defined by \eqref{BExpr},
in general $B_{36}$, $B_{35}$, $B_{34}$ are polynomials of degree $3N$,
$B_{26}$, $B_{25}$, $B_{24}$ are polynomials of degree $2N$,
and $B_{16}(\lambda)\,{=}\,\gamma_6(\lambda)$, $B_{15}(\lambda)\,{=}\,\gamma_5(\lambda)$,
$B_{14}(\lambda)\,{=}\,\gamma_4(\lambda)$. We denote by $\tens{B}(\lambda_k)$ the
matrix polynomial serving as a coefficient matrix of \eqref{Minors23}, namely:
$$\tens{B}(\lambda) = \tens{B}_{3N} \lambda^{3N} + \dots + \tens{B}_1 \lambda + \tens{B}_0.$$

The system of equations \eqref{Minors23} is the main result of the proposed scheme.
It allows to compute the set of points $\{(\lambda_k,w_k)\}$ serving as variables of separation
for Hamiltonian systems on a generic orbit of SU(4) loop group.
In what follows we consider \eqref{Minors23} as equations for $(\lambda,w)$:
\begin{equation}\label{LWeqs}
\tens{B}(\lambda) \begin{bmatrix} w^2 \\ w \\ 1 \end{bmatrix} = \bm{0}.
\end{equation}
Nontrivial solutions for $w$ exist if
\begin{equation}\label{ConsistentEqNLS}
\det \tens{B}(\lambda) =0,
\end{equation}
which is an algebraic equation of degree $6N$ if $\tens{B}_{3N}$ is nonsingular.
Roots of $\det \tens{B}$ give the set of $\lambda$-variables
forming a half of variables of separation, we suppose all $\lambda_k$ are distinct.
At every $\lambda_k$ the $\tens{L}$-matrix has 4 eigenvalues,
one on every sheet of the spectral curve~$\mathcal{R}$.

We solve \eqref{LWeqs} by the Gauss method:
\begin{gather}
\begin{bmatrix} 0 & \scriptsize -\frac{1}{B_{14}}\begin{vmatrix} B_{16}(\lambda) & B_{26}(\lambda)\\
 B_{14}(\lambda) & B_{24}(\lambda) \end{vmatrix} &
 \scriptsize -\frac{1}{B_{14}}\begin{vmatrix} B_{16}(\lambda) & B_{36}(\lambda)\\
 B_{14}(\lambda) & B_{34}(\lambda) \end{vmatrix} \\
0 & \scriptsize -\frac{1}{B_{14}} \begin{vmatrix} B_{15}(\lambda) & B_{25}(\lambda)\\
 B_{14}(\lambda) & B_{24}(\lambda) \end{vmatrix} &
 \scriptsize -\frac{1}{B_{14}} \begin{vmatrix} B_{15}(\lambda) & B_{35}(\lambda)\\
 B_{14}(\lambda) & B_{34}(\lambda) \end{vmatrix} \\
B_{14}(\lambda) & B_{24}(\lambda) & B_{34}(\lambda) \end{bmatrix}
\begin{bmatrix} w^2 \\ w \\ 1 \end{bmatrix} = \bm{0},\notag \\
w = -\frac{\scriptsize \begin{vmatrix} B_{15}(\lambda) & B_{35}(\lambda)\\
 B_{14}(\lambda) & B_{34}(\lambda) \end{vmatrix}}
 {\scriptsize \begin{vmatrix} B_{15}(\lambda) & B_{25}(\lambda)\\
 B_{14}(\lambda) & B_{24}(\lambda) \end{vmatrix}}=
 -\frac{\scriptsize \begin{vmatrix} B_{16}(\lambda) & B_{36}(\lambda)\\
 B_{14}(\lambda) & B_{34}(\lambda) \end{vmatrix}}
 {\scriptsize \begin{vmatrix} B_{16}(\lambda) & B_{26}(\lambda)\\
 B_{14}(\lambda) & B_{24}(\lambda) \end{vmatrix}} =
 -\frac{\scriptsize \begin{vmatrix} B_{16}(\lambda) & B_{36}(\lambda)\\
 B_{15}(\lambda) & B_{35}(\lambda) \end{vmatrix}}
 {\scriptsize \begin{vmatrix} B_{16}(\lambda) & B_{26}(\lambda)\\
 B_{15}(\lambda) & B_{25}(\lambda) \end{vmatrix}}. \label{GaussSol}
\end{gather}
The last expression for $w$ is obtained by using row 2 as leading. In this way we compute
one $w_k$ for each $\lambda_k$, and all the points $(\lambda_k,\,w_k)$ are located
on the same sheet of the spectral curve~$\mathcal{R}$ defined by \eqref{SpectrCurve}.

The second matrix equation \eqref{RootEqNLS} gives
\begin{gather}
w_k^4 = w_k^2 a_2(\lambda_k) + w_k a_3(\lambda_k) + a_4(\lambda_k),\quad \text{or} \notag\\
\big[w_k + \alpha_3(\lambda_k)\big]  \big[w_k^3 - \alpha_3(\lambda_k) w_k^2
- A_2(\lambda_k) w_k - A_3(\lambda_k) \big] =0, \label{SpectCurveRed}
\end{gather}
where $a_2$, $a_3$, $a_4$, $A_2$, $A_3$ are polynomials with the coefficients defined by \eqref{AExpr}.
The equation \eqref{SpectCurveRed} gives a simplification of the spectral curve equation \eqref{SpectrCurve}
realized at every point of the set $\{\vec{\lambda},\,\vec{w}\}\,{\equiv}\,\{(\lambda_k,\, w_k)\,{:}$ $k\,{=}\,1,\,
\dots,\, 6 N\}$.
The variables $\{\vec{\lambda},\,\vec{w}\}$ give another parametrization of $\mathcal{O}_\text{f}$,
and serve as variables of separation (see a proof in \cite{Adams}).

\subsection{Connection to the known results}
Recall some known results. In \cite{Sklyanin92} one can find Sklyanin's Conjecture\;1
asserting an existence of a polynomial~$\mathcal{B}$ whose roots serve as a half of variables of separation
and a function $\mathcal{A}$ giving the other half of variables of separation, that is
\begin{equation*}
  \mathcal{B}(\lambda_k)=0,\qquad
  w_k = \mathcal{A}(\lambda_k),
\end{equation*}
where $\{(\lambda_k,w_k)\}$ are canonically conjugate:
\begin{equation*}
  \{\lambda_k,\lambda_l\} =0, \qquad
  \{\lambda_k, w_l\} = \delta_{kl}, \qquad
  \{w_k,w_l\}=0.
\end{equation*}
In \cite{Sklyanin92} explicit expressions for the classical SL(3) magnetic chain are presented,
and canonical conjugation is proven for this particular case. A fine proof of canonical conjugation
for the loop group GL($r$) of arbitrary rank $r$ can be found in \cite{Adams}.

A thorough development of Sklyanin's idea is realized by Gekhtman in \cite{Gekhtman}, where
formulas for calculation of $\mathcal{B}$ and $\mathcal{A}$ are presented.
Unfortunately, in \cite{Gekhtman} the result is given as a calculation technique
without any explanation of grounds. Such situation provokes further investigation of the problem.
One of explanation based on orbit method is presented in \cite{BernHol2013}.
Its development for the SL(4) case is presented in this paper.

Regarding Conjecture\;1 from Sklyanin's paper  the polynomial $\det \tens{B}$
is the polynomial~$\mathcal{B}$ whose roots serve as a half of variables of separation.
And the expressions \eqref{GaussSol} serve as the function $\mathcal{A}$,
compare them with the result from \cite{Gekhtman}.
After Gekhtman one should take the last column without last entry of $\tens{L}$-matrix, we denote
this vector by $\bm{\xi}$, and the rest of $\tens{L}$-matrix without the last row, we denote this matrix by $\tens{T}$.
To match this computation with the above expressions we use the transposed $\tens{L}$-matrix
(without loss of generality):
$$\bm{\xi} = \small \begin{bmatrix} \gamma_6(\lambda) \\ \gamma_5(\lambda) \\ \gamma_4(\lambda) \end{bmatrix},\qquad
\tens{T} = \begin{bmatrix} \alpha_1(\lambda) & \gamma_1(\lambda) & \gamma_3(\lambda) \\
\beta_1(\lambda) & \alpha_1(\lambda)-\alpha_1(\lambda) & \gamma_2(\lambda) \\
\beta_3(\lambda) & \beta_2(\lambda) & \alpha_3(\lambda)-\alpha_2(\lambda) \end{bmatrix}.
$$
Then construct the square matrices:
$$\tens{S}\,{=}\,[\bm{\xi},\,\tens{T}\bm{\xi},\,\tens{T}^{2}\bm{\xi}],\quad
\tens{S}^{(1)}\,{=}\,[\bm{t}^{(1)},\,\bm{\xi},\,\tens{T}\bm{\xi}],\quad
\tens{S}^{(2)}\,{=}\,[\bm{t}^{(2)},\,\bm{\xi},\,\tens{T}\bm{\xi}],\quad
\tens{S}^{(3)}\,{=}\,[\bm{t}^{(3)},\,\bm{\xi},\,\tens{T}\bm{\xi}],$$
where $\bm{t}^{(j)}$ are columns of $\tens{T}$.
Then $\mathcal{B}$ and $\mathcal{A}$ are given by the
formulas:
\begin{equation}\label{ABGekhtman}
\mathcal{B}(\lambda) = \det \tens{S},\qquad
\mathcal{A}(\lambda) = \frac{\det \tens{S}^{(1)}}{\det \check{\tens{S}}_1^3} =
-\frac{\det \tens{S}^{(2)}}{\det \check{\tens{S}}_2^3} =
\frac{\det \tens{S}^{(3)}}{\det \check{\tens{S}}_3^3},
\end{equation}
where $\check{\tens{S}}_{k}^{j}$ is obtained from $\tens{S}$ by elimination of row $k$ and column $j$.

Expressions for $\mathcal{B}$ and $\mathcal{A}$ functions given by
\eqref{GaussSol} and \eqref{ABGekhtman} coincide. The matrices $\tens{B}$ and $\tens{S}$ differ in a singular matrix,
the same is true for the other corresponding matrices in the expressions:
\begin{gather*}
{\scriptsize\begin{vmatrix} B_{15}(\lambda) & B_{35}(\lambda)\\
 B_{14}(\lambda) & B_{34}(\lambda) \end{vmatrix}} = - \det \tens{S}^{(1)},\quad
{\scriptsize\begin{vmatrix} B_{16}(\lambda) & B_{36}(\lambda)\\
 B_{14}(\lambda) & B_{34}(\lambda) \end{vmatrix}} = \det \tens{S}^{(2)},\quad
{\scriptsize\begin{vmatrix} B_{16}(\lambda) & B_{36}(\lambda)\\
 B_{15}(\lambda) & B_{35}(\lambda) \end{vmatrix}} = - \det \tens{S}^{(3)},\\
{\scriptsize\begin{vmatrix} B_{15}(\lambda) & B_{25}(\lambda)\\
 B_{14}(\lambda) & B_{24}(\lambda) \end{vmatrix}} = \det \check{\tens{S}}_1^3,\qquad
{\scriptsize\begin{vmatrix} B_{16}(\lambda) & B_{26}(\lambda)\\
 B_{14}(\lambda) & B_{24}(\lambda) \end{vmatrix}} = \det \check{\tens{S}}_2^3, \qquad
 {\scriptsize\begin{vmatrix} B_{16}(\lambda) & B_{26}(\lambda)\\
 B_{15}(\lambda) & B_{25}(\lambda) \end{vmatrix}} = \det \check{\tens{S}}_3^3.
\end{gather*}

The problem of separation of variables on coadjoint orbits of the loop group GL($r$)
is also considered in \cite{Adams,AdamsCO}.
For further explanation we introduce the matrix  $\tens{N}(\lambda,w)\,{\equiv}\,\tens{L}(\lambda)\,{-}\,w\Ibb$,
and denote by $\widetilde{\tens{N}}$ its adjoint matrix
whose entries $\widetilde{N}_{ij}$ are cofactors corresponding to the entries $N_{ji}$ of $\tens{N}$.
Adams, Harnad and Hurtubise show that variables of separation (spectral Darboux coordinates)
are zeros of $\widetilde{\tens{N}}(\lambda,w)\bm{v}_0$ with an arbitrary vector $\bm{v}_0$ usually chosen as $(1,\,0,\,\dots,\,0)^\text{T}$. Applying this idea to the orbit $\mathcal{O}_{\text{f}}$ in the loop algebra
$\mathfrak{sl}(4)$ we replace $\widetilde{\tens{N}}$ by its transpose and use another vector $\bm{v}_0$
$$\widetilde{\tens{N}}^{\text{t}}(\lambda,w) \small \begin{pmatrix} 0 \\ 0 \\ 0 \\ 1 \end{pmatrix} =
\begin{pmatrix} \widetilde{N}_{41}(\lambda,w) \\ \widetilde{N}_{42}(\lambda,w) \\
\widetilde{N}_{43}(\lambda,w) \\ \widetilde{N}_{44}(\lambda,w)  \end{pmatrix} = \bm{0}.$$
\begin{gather*}
\begin{split}
&\widetilde{N}_{41} (\lambda,w)  \equiv \scriptsize \begin{vmatrix}
\gamma_1(\lambda) & \alpha_2(\lambda) - \alpha_1(\lambda) - w & \beta_2(\lambda) \\
\gamma_3(\lambda) & \gamma_2(\lambda) & \alpha_3(\lambda) - \alpha_2(\lambda) - w \\
\gamma_6(\lambda) & \gamma_5(\lambda) & \gamma_4(\lambda)
\end{vmatrix} =  w^2 B_{16}(\lambda) + w B_{26}(\lambda) + B_{36},\\
&\widetilde{N}_{42} (\lambda,w)  \equiv  \scriptsize - \begin{vmatrix}
\alpha_1(\lambda) - w & \beta_1(\lambda) & \beta_3 (\lambda)\\
\gamma_3(\lambda) & \gamma_2(\lambda) & \alpha_3(\lambda) - \alpha_2(\lambda) - w \\
\gamma_6(\lambda) & \gamma_5(\lambda) & \gamma_4(\lambda)
\end{vmatrix} = w^2 B_{15}(\lambda) + w B_{25}(\lambda) + B_{35},\\
&\widetilde{N}_{43} (\lambda,w)  \equiv \scriptsize - \begin{vmatrix}
\alpha_1(\lambda) - w & \beta_1(\lambda) & \beta_3 (\lambda)\\
\gamma_1(\lambda) & \alpha_2(\lambda) - \alpha_1(\lambda) - w & \beta_2(\lambda) \\
\gamma_6(\lambda) & \gamma_5(\lambda) & \gamma_4(\lambda)
\end{vmatrix} = w^2 B_{14}(\lambda) + w B_{24}(\lambda) + B_{34},
\end{split}
\end{gather*}
One can see that \eqref{LWeqs} coincides with the following
\begin{gather}\label{MinorEqs}
\widetilde{N}_{41} (\lambda,w) = 0,\qquad
\widetilde{N}_{42} (\lambda,w) = 0,\qquad
\widetilde{N}_{43} (\lambda,w) = 0.
\end{gather}
The last equation is a simplification of the spectral curve equation, true only for the set $\{(\lambda_k,w_k)\}$
satisfying both the other two equations. That is why the equation $\widetilde{N}_{44}(\lambda,w)\,{=}\,0$.

We see the obtained result are in good correspondence with the known ones. Moreover,
the proposed orbit approach gives an obvious geometric explanation
to the algorithm declared in \cite{Sklyanin95}, and can be easily extended to algebras of higher rank.

\subsection{Mnemonic rule and algorithm of calculation}
Using a mnemonic rule we obtain
\begin{gather*}
\begin{pmatrix} I_2(\lambda) \\ I_3(\lambda) \\ I_4(\lambda) \end{pmatrix}  =
\tens{B}^{\text{t}}(\lambda)
\begin{pmatrix} \beta_6(\lambda) \\ \beta_5(\lambda) \\ \beta_4(\lambda) \end{pmatrix} +
\begin{pmatrix}  A_2(\lambda) + \alpha_3^2(\lambda)  \\
A_3(\lambda) + \alpha_3(\lambda) A_2(\lambda)\\ \alpha_3(\lambda) A_3(\lambda)  \end{pmatrix}\\
\tens{B}(\lambda) = \small \begin{bmatrix} \gamma_6 &
  \begin{vmatrix} \gamma_1 & \alpha_2-\alpha_1 \\ \gamma_6 & \gamma_5 \end{vmatrix} +
  \begin{vmatrix} \gamma_3 & \alpha_3-\alpha_2 \\ \gamma_6 & \gamma_4 \end{vmatrix} &
  \begin{vmatrix} \gamma_1 & \alpha_2 - \alpha_1 & \beta_2 \\
  \gamma_3 & \gamma_2 & \alpha_3 - \alpha_2 \\
  \gamma_6 & \gamma_5 & \gamma_4 \end{vmatrix} \\
  \gamma_5 & -\begin{vmatrix} \alpha_1 & \beta_1 \\ \gamma_6 & \gamma_5 \end{vmatrix} +
  \begin{vmatrix} \gamma_2 & \alpha_3-\alpha_2 \\ \gamma_5 & \gamma_4 \end{vmatrix} &
  -\begin{vmatrix} \alpha_1 & \beta_1 & \beta_3 \\
  \gamma_3 & \gamma_2 & \alpha_3 - \alpha_2 \\
  \gamma_6 & \gamma_5 & \gamma_4  \end{vmatrix} \\
  \gamma_4 & -\begin{vmatrix} \alpha_1 & \beta_3 \\ \gamma_6 & \gamma_4 \end{vmatrix} -
  \begin{vmatrix} \alpha_2-\alpha_1 & \beta_2 \\ \gamma_5 & \gamma_4 \end{vmatrix} &
  \begin{vmatrix} \alpha_1 & \beta_1 & \beta_3 \\
  \gamma_1 & \alpha_2 - \alpha_1 & \beta_2 \\
  \gamma_6 & \gamma_5 & \gamma_4  \end{vmatrix}
  \end{bmatrix}(\lambda),\\
\begin{split}
&A_2(\lambda) = \alpha_1^2 + \alpha_2^2 -
  \alpha_1\alpha_2 - \alpha_2\alpha_3 + \beta_1\gamma_1 + \beta_2\gamma_2 + \beta_3\gamma_3 = \\
  &\phantom{A_2(\lambda)} = \small -\begin{vmatrix} \alpha_1 & \beta_1   \\
  \gamma_1 & \alpha_2-\alpha_1 \end{vmatrix} -
  \begin{vmatrix} \alpha_2-\alpha_1 & \beta_2  \\
  \gamma_2 & \alpha_3-\alpha_2 \end{vmatrix} -
  \begin{vmatrix} \alpha_1 & \beta_3  \\
  \gamma_3 & \alpha_3-\alpha_2 \end{vmatrix}. \\
&A_3(\lambda)= \alpha_2\alpha_1^2 -
  \alpha_1 \alpha_2^2 + \alpha_1\alpha_2\alpha_3 - \alpha_3 \alpha_1^2 -
  \beta_1\gamma_1(\alpha_3-\alpha_2) - \beta_2\gamma_2\alpha_1 - \beta_3\gamma_3(\alpha_2-\alpha_1) +\\
  &\phantom{A_3(\lambda)} + \beta_3 \gamma_1 \gamma_2 + \beta_1 \beta_2 \gamma_3 =
  \small \begin{vmatrix} \alpha_1 & \beta_1 & \beta_3 \\
  \gamma_1 & \alpha_2-\alpha_1 & \beta_2 \\
  \gamma_3 & \gamma_2 & \alpha_3-\alpha_2 \end{vmatrix}.
\end{split}
\end{gather*}
One can calculate the $\tens{B}$-matrix in the following way:
$$\tens{B}(\lambda) = \begin{bmatrix}
\frac{\partial I_2(\lambda)}{\partial \beta_6^{(0)}}
& \frac{\partial I_3(\lambda)}{\partial \beta_6^{(0)}}
& \frac{\partial I_4(\lambda)}{\partial \beta_6^{(0)}} \\
\frac{\partial I_2(\lambda)}{\partial \beta_5^{(0)}}
& \frac{\partial I_3(\lambda)}{\partial \beta_5^{(0)}}
& \frac{\partial I_4(\lambda)}{\partial \beta_5^{(0)}} \\
\frac{\partial I_2(\lambda)}{\partial \beta_4^{(0)}}
& \frac{\partial I_3(\lambda)}{\partial \beta_4^{(0)}}
& \frac{\partial I_4(\lambda)}{\partial \beta_4^{(0)}}
\end{bmatrix}.$$
Also $A_k$ is the sum of diagonal $k^{\text{th}}$ minors of the left upper $3\,{\times}\,3$ block of $\tens{L}$.

We are looking for special points $\{(\lambda_k,w_k)\}$ where
\begin{equation*}
\big[ \beta_6(\lambda), \beta_5(\lambda), \beta_4(\lambda) \big]
\tens{B}(\lambda) \begin{bmatrix} w^2 \\ w \\ 1 \end{bmatrix}= 0
\end{equation*}
for any $\beta_6^{(m)}$, $\beta_5^{(m)}$, $\beta_4^{(m)}$, $m\,{=}\,0,\,\dots \mathcal{N}\,{-}\,1$.
This requirement leads to the equations \eqref{LWeqs}, which is the zero-eigenvalue problem for $\tens{B}$.
Nontrivial solutions exist if $\tens{B}$ is singular, that is realized at $6N$ values of $\lambda$, we denote them $\{\lambda_k\}_{k=1}^{6N}$ and suppose all distinct.
Note, at every $\lambda_k$ the spectral curve
\begin{equation*}
w^4 = \big[I_2(\lambda), I_3(\lambda), I_4(\lambda) \big]
\begin{bmatrix} w^2 \\ w \\ 1 \end{bmatrix}
\end{equation*}
is reduced to a simpler form:
\begin{equation*}
w^4 = \big(w + \alpha_3(\lambda_k)\big)
\big(w^3 - \alpha_3(\lambda_k) w^2 - A_2(\lambda_k) w - A_3(\lambda_k) \big).
\end{equation*}

Then for $\tens{B}(\lambda_k)$ we need an eigenvector of
the form $\big[w^2,\,w,\,1\big]$ corresponding to the zero eigenvalue --- this is the mathematical sense of
\eqref{LWeqs}, see also \cite{Adams}. This requirement is realizable, because entries of $\tens{B}$
are not predefined, but expressed through dynamical variables. It turns out the equations \eqref{GaussSol}
guarantee a one-to-one map between dynamic and spectral variables.

A real process of solution starts from a set of points $\{(\lambda_k,w_k)\}$ on the spectral curve,
then according from \eqref{Minors23} one can obtain unambiguous expressions for dynamic variables
in terms of the spectral variables.

\section{Separation of variables Theorems}
Summarizing the above computation we formulate the following
\begin{SoV}\label{T:SoVNLS}
Suppose the orbit $\mathcal{O}_{\text{f}}$ is parameterized by the variables
 $\{\alpha_1^{(m)}$, $\alpha_2^{(m)}$, $\alpha_3^{(m)}$, $\beta_1^{(m)}$, $\beta_2^{(m)}$, $\beta_3^{(m)}$,
 $\gamma_1^{(m)}$, $\gamma_2^{(m)}$, $\gamma_3^{(m)}$, $\gamma_4^{(m)}$, $\gamma_5^{(m)}$,
 $\gamma_6^{(m)}\,{:}$ $m\,{=}\,0,\,\dots,\, N\,{-}\,1\}$ as above.
Then the new variables $\{(\lambda_k, w_k)\,{:}$ $k\,{=}\,1,\,\dots,\,6 N\}$
defined by the formulas
\begin{equation}\label{newvarNLS}
  \mathcal{B}(\lambda_k)=0,\qquad
  w_k = \mathcal{A}(\lambda_k),
\end{equation}
where $\mathcal{B}$ is the polynomial of degree $6 N$ and $\mathcal{A}$
is the algebraic function such that
\begin{subequations}\label{ABNLS}
\begin{align}
\mathcal{B}(\lambda) &= \det \tens{B}(\lambda),
  \label{NilpotentPolyNLS}\\   \label{w1PolyNLS}
  \mathcal{A}(\lambda) & = -\frac{\scriptsize \begin{vmatrix} B_{15}(\lambda) & B_{35}(\lambda)\\
 B_{14}(\lambda) & B_{34}(\lambda) \end{vmatrix}}
 {\scriptsize \begin{vmatrix} B_{15}(\lambda) & B_{25}(\lambda)\\
 B_{14}(\lambda) & B_{24}(\lambda) \end{vmatrix}}\quad \text{or} \quad
 -\frac{\scriptsize \begin{vmatrix} B_{16}(\lambda) & B_{36}(\lambda)\\
 B_{14}(\lambda) & B_{34}(\lambda) \end{vmatrix}}
 {\scriptsize \begin{vmatrix} B_{16}(\lambda) & B_{26}(\lambda)\\
 B_{14}(\lambda) & B_{24}(\lambda) \end{vmatrix}}\quad \text{or} \quad
 -\frac{\scriptsize \begin{vmatrix} B_{16}(\lambda) & B_{36}(\lambda)\\
 B_{15}(\lambda) & B_{35}(\lambda) \end{vmatrix}}
 {\scriptsize \begin{vmatrix} B_{16}(\lambda) & B_{26}(\lambda)\\
 B_{15}(\lambda) & B_{25}(\lambda) \end{vmatrix}}
\end{align}
\end{subequations}
have the following properties:
\begin{enumerate}
\item[\textup{(i)}]  a pair $(\lambda_k,w_k)$ is a root of the characteristic
polynomial \eqref{CharPoly};
\item[\textup{(ii)}] a pair
$(\lambda_k, w_k)$ is canonically conjugate with respect to
the first Lie-Poisson bracket \eqref{LiePoissonBraNLS}:
\begin{equation}\label{PoissonBraNLSConj}
  \{\lambda_k,\lambda_l\}_{\textup{f}} =0, \qquad
  \{\lambda_k, w_l\}_{\textup{f}} = \delta_{kl}, \qquad
  \{w_k,w_l\}_{\textup{f}}=0;
\end{equation}
\item[\textup{(iii)}] the corresponding  Liouville 1-form is
\begin{gather*}
\Omega_{\textup{f}}=\sum\limits_{k} w_{k}\,d\lambda_{k}.
\end{gather*}
\end{enumerate}
\end{SoV}

\begin{proof}
(i) The characteristic polynomial $P$ defined by \eqref{CharPoly} has
$\mathcal{B}(\lambda_k)$ as a factor at every point $(\lambda_k,w_k)$, that is proven by explicit calculation.
All the expressions \eqref{w1PolyNLS} for $\mathcal{A}$ coincide at all zeros
$\{\lambda_k\}$ of $\mathcal{B}$, that is they give the same eigenvalue of the $\tens{L}$-matrix \eqref{muExpr}.
This proves the assertion (i).

(ii) To prove this assertion we use the Conjugate variable lemma\;1 proven in \cite{BernHol2013} and recalled here,
and $\mathcal{A}$-$\mathcal{B}$ bracket lemma\;1 stated below.

\begin{ConjVLemma}\label{L:ConjVarNLS}
If $\mathcal{B}$ and $\mathcal{A}$ satisfy the following identities with respect to
the first Lie-Poisson bra\-cket \eqref{PoissonBraNLS}
$$\{\mathcal{B}(u),\mathcal{B}(v)\}_{\textup{f}} = 0,\quad
\{\mathcal{A}(u),\mathcal{A}(v)\}_{\textup{f}} = 0, \quad
\{\mathcal{A}(u),\mathcal{B}(v)\}_{\textup{f}} = \frac{f(u,v)\mathcal{B}(u)-\mathcal{B}(v)}{u-v},$$
where $f$ is an arbitrary function such that $\lim_{v\to u} f(u,v)\,{=}\,1$,
then the variables $\{(\lambda_k,w_k)\,{:}$ $k\,{=}\,1,\,\dots,\,6 N\}$ defined by
$$\mathcal{B}(\lambda_k)=0,\qquad w_k=\mathcal{A}(\lambda_k)$$
are canonically conjugate with respect to $\{\cdot,\cdot\}_{\textup{f}}$:
$$\{\lambda_k,\lambda_l\}_{\textup{f}} = 0,\qquad
\{\lambda_k,w_l\}_{\textup{f}} = \delta_{kl},\qquad \{w_k,w_l\}_{\textup{f}} = 0. $$
\end{ConjVLemma}

\begin{ABLemma}\label{L:ABbracketNLS}
For $\mathcal{B}$ and $\mathcal{A}$ defined by \eqref{ABNLS} the following identities are true
with respect to the first Lie-Poisson bra\-cket \eqref{PoissonBraNLS}:
$$\{\mathcal{B}(u),\mathcal{B}(v)\}_{\textup{f}} = 0,\quad
\{\mathcal{A}(u),\mathcal{A}(v)\}_{\textup{f}} = 0, \quad
\{\mathcal{A}(u),\mathcal{B}(v)\}_{\textup{f}} = \frac{f(u,v)\mathcal{B}(u)-\mathcal{B}(v)}{u-v},$$
where
$$f(u,v)=\Bigg(\frac{\scriptsize \begin{vmatrix} B_{15}(v) & B_{25}(v)\\
 B_{14}(v) & B_{24}(v) \end{vmatrix}}
 {\scriptsize \begin{vmatrix} B_{15}(u) & B_{25}(u)\\
 B_{14}(u) & B_{24}(u) \end{vmatrix}}\Bigg)^2\quad \text{or} \quad
 \Bigg(\frac{\scriptsize \begin{vmatrix} B_{16}(v) & B_{26}(v)\\
 B_{14}(v) & B_{24}(v) \end{vmatrix}}
 {\scriptsize \begin{vmatrix} B_{16}(u) & B_{26}(u)\\
 B_{14}(u) & B_{24}(u) \end{vmatrix}}\Bigg)^2\quad \text{or} \quad
 \Bigg(\frac{\scriptsize \begin{vmatrix} B_{16}(v) & B_{26}(v)\\
 B_{15}(v) & B_{25}(v) \end{vmatrix}}
 {\scriptsize \begin{vmatrix} B_{16}(u) & B_{26}(u)\\
 B_{15}(u) & B_{25}(u) \end{vmatrix}}\Bigg)^2\quad$$
for the $\mathcal{A}$-functions from \eqref{w1PolyNLS}, respectively.
\end{ABLemma}

(iii) The Liouville 1-form on $\mathcal{O}_{\text{f}}$ is
implied by (\ref{PoissonBraNLSConj}):
\begin{equation*}
\Omega_{\text{f}}=\sum\limits_{k} w_{k}\,d\lambda_{k}.
\end{equation*}
By fixing values of the Hamiltonians $h_2^{(0)}$, $h_2^{(1)}$, \ldots, $h_2^{(N-1)}$,
$h_3^{(0)}$, $h_3^{(1)}$, \ldots, $h_3^{(2N-1)}$, $h_4^{(0)}$, $h_4^{(1)}$, \ldots, $h_4^{(3N-1)}$
we obtain a Liouville torus.
On the torus every variable $w_k$ becomes an algebraic function of the corresponding conjugate variable $\lambda_k$
due to (\ref{HyperCurveNLS}),
and the form $\Omega_{\text{f}}$ becomes a sum of
meromorphic differentials on the Riemann surface $P(w,\lambda)=0$.

This completes the proof of \nameSoV\;\ref{T:SoVNLS}.
\end{proof}

Given a set of pairs $\{(\lambda_k, w_k)\,{:}\,k\,{=}\,1,\,\dots,\, 6 N\}$ we are able to compute
the dynamic variables satisfying the equations (\ref{newvarNLS}). Thus, one defines a homomorphism
\begin{equation}\label{homomorfizmNLS}
  \Complex^{6 N} \to \mathcal{O}_\text{f}
\end{equation}
that maps $\{(\lambda_k, w_k)\}$ to a point of $\mathcal{O}_\text{f}$.
When all the Hamiltonians are fixed the homomorphism \eqref{homomorfizmNLS} turns into the
map from the symmetrized product of $6N$ Riemann surfaces $\mathcal{R}$
defined by \eqref{HyperCurveNLS} to a Liouville torus:
\begin{equation*}
\Sym\{\mathcal{R}\times \mathcal{R}\times \cdots \times
\mathcal{R}\} \mapsto T^{6N}.
\end{equation*}

\subsection*{Separation of variables on $\mathcal{O}_{\text{s}}$}

Separation of variables on $\mathcal{O}_{\text{s}}$ is also realized through
restriction to the orbit, see \cite{BernHol2013} for details.
One obtains the same matrix equation \eqref{Minors23} producing
the same expressions for $\mathcal{A}$ and $\mathcal{B}$. It means the same points
on the spectral curve \eqref{SpectrCurve} serve as variables of separation.

\begin{SoV}\label{T:SoVHM}
Suppose the orbit $\mathcal{O}_{\text{s}}$ is parameterized by the variables
 $\{\alpha_1^{(m)}$, $\alpha_2^{(m)}$, $\alpha_3^{(m)}$, $\beta_1^{(m)}$, $\beta_2^{(m)}$, $\beta_3^{(m)}$,
 $\gamma_1^{(m)}$, $\gamma_2^{(m)}$, $\gamma_3^{(m)}$, $\gamma_4^{(m)}$, $\gamma_5^{(m)}$,
 $\gamma_6^{(m)}\,{:}$ $m\,{=}\,0,\,\dots,\, N\,{-}\,1\}$ as above.
Then the new variables $\{(\lambda_k, w_k)\,{:}$ $k\,{=}\,1,\,\dots,\,6 N\}$
defined by the formulas \eqref{newvarNLS}, \eqref{ABNLS}
have the following properties
\begin{enumerate}
\item[\textup{(i)}]  a pair $(\lambda_k,w_k)$ is a root of the characteristic
polynomial \eqref{CharPoly}.
\item[\textup{(ii)}] a pair
$(\lambda_k, w_k)$ is quasi-canonically conjugate with respect to
the second Lie-Poisson bracket \eqref{LiePoissonBraHM}:
\begin{equation}\label{PoissonBraHMConj}
  \{\lambda_k,\lambda_l\}_{\textup{s}} =0 \qquad
  \{\lambda_k, w_l\}_{\textup{s}} = -\lambda_k^{ N}\delta_{kl}, \qquad
  \{w_k,w_l\}_{\textup{s}}=0;
\end{equation}
\item[\textup{(iii)}] the corresponding  Liouville 1-form is
\begin{gather*}
\Omega_{\textup{s}}= - \sum\limits_{k} \lambda_k^{- N} w_{k}\,d\lambda_{k}.
\end{gather*}
\end{enumerate}
\end{SoV}

\begin{proof}
(i) The proof repeats one for the \nameSoV\,\ref{T:SoVNLS}.

(ii) The assertion follows from the lemmas below.
\begin{ConjVLemma}\label{L:ConjVarHM}
If $\mathcal{B}$ and $\mathcal{A}$ satisfy the following identities
with respect to the second Lie-Poisson bracket \eqref{LiePoissonBraHM}
\begin{gather*}
\{\mathcal{B}(u),\mathcal{B}(v)\}_{\textup{s}} = 0,\quad
\{\mathcal{A}(u),\mathcal{A}(v)\}_{\textup{s}} = 0, \\
\{\mathcal{A}(u),\mathcal{B}(v)\}_{\textup{s}} =
\frac{u^{ N}\mathcal{B}(v)-v^{ N}\mathcal{B}(u) f(u,v)}{u-v},
\end{gather*}
where $f$ is an arbitrary function such that $\lim_{v\to u} f(u,v)\,{=}\,1$,
then the variables $\{(\lambda_k,w_k)\}$ defined by
$$\mathcal{B}(\lambda_k)=0,\qquad w_k=\mathcal{A}(\lambda_k)$$
are quasi-canonically conjugate with respect to $\{\cdot,\cdot\}_{\textup{s}}$:
$$\{\lambda_k,\lambda_l\}_{\textup{s}} = 0,\qquad \{\lambda_k,w_l\}_{\textup{s}} =
-\lambda_k^{ N}\delta_{kl},\qquad \{w_k,w_l\}_{\textup{s}} = 0.$$
\end{ConjVLemma}

\begin{ABLemma}\label{L:ABbracketHM}
For $\mathcal{B}$ and $\mathcal{A}$ defined by \eqref{ABNLS} the following identities are true
with respect to the second Lie-Poisson bracket \eqref{LiePoissonBraHM}
\begin{gather*}
\{\mathcal{B}(u),\mathcal{B}(v)\}_{\textup{s}} = 0,\quad
\{\mathcal{A}(u),\mathcal{A}(v)\}_{\textup{s}} = 0, \\
\{\mathcal{A}(u),\mathcal{B}(v)\}_{\textup{s}} =
\frac{u^{ N}\mathcal{B}(v)-v^{ N}\mathcal{B}(u) f(u,v)}{u-v},
\end{gather*}
where
$$f(u,v)=\Bigg(\frac{\scriptsize \begin{vmatrix} B_{15}(v) & B_{25}(v)\\
 B_{14}(v) & B_{24}(v) \end{vmatrix}}
 {\scriptsize \begin{vmatrix} B_{15}(u) & B_{25}(u)\\
 B_{14}(u) & B_{24}(u) \end{vmatrix}}\Bigg)^2\quad \text{or} \quad
 \Bigg(\frac{\scriptsize \begin{vmatrix} B_{16}(v) & B_{26}(v)\\
 B_{14}(v) & B_{24}(v) \end{vmatrix}}
 {\scriptsize \begin{vmatrix} B_{16}(u) & B_{26}(u)\\
 B_{14}(u) & B_{24}(u) \end{vmatrix}}\Bigg)^2\quad \text{or} \quad
 \Bigg(\frac{\scriptsize \begin{vmatrix} B_{16}(v) & B_{26}(v)\\
 B_{15}(v) & B_{25}(v) \end{vmatrix}}
 {\scriptsize \begin{vmatrix} B_{16}(u) & B_{26}(u)\\
 B_{15}(u) & B_{25}(u) \end{vmatrix}}\Bigg)^2\quad$$
for the $\mathcal{A}$-functions from \eqref{w1PolyNLS}, respectively.
\end{ABLemma}

(iii) The Liouville 1-form on $\mathcal{O}_{\text{s}}$ is implied by (\ref{PoissonBraHMConj}):
\begin{equation*}
\Omega_{\text{s}}=-\sum\limits_{k} \lambda^{- N} w_{k}\,d\lambda_{k}.
\end{equation*}
Reduction to a Liouville torus is realized by fixing values of the
Hamiltonians $h_2^{(N)}$, $h_2^{(N+1)}$, \ldots, $h_2^{(2N-1)}$,
$h_3^{(N)}$, $h_3^{(N+1)}$, \ldots, $h_3^{(3N-1)}$, $h_4^{(N)}$, $h_4^{(N+1)}$, \ldots, $h_4^{(4N-1)}$.
On the torus every $w_k$
is an algebraic function of the conjugate variable $\lambda_k$.
After this reduction the form $\Omega_{\text{s}}$ becomes a sum of
meromorphic differentials on the Riemann surface $P(w,\lambda)=0$.

This completes the proof of \nameSoV\;\ref{T:SoVHM}.
\end{proof}

Above we suppose the matrix polynomial~$\tens{B}$ has the maximal degree $6N$. If not
one should apply to $\tens{L}$-matrix a proper similarity transformation that makes
$\tens{B}$ of maximal degree.

\section{Conclusion and Discussion}
Here a brief summary of the orbit approach is given.
Recall that an integrable system is constructed on a coadjoint orbit
in the loop Lie algebra $\widetilde{\mathfrak{g}}$, it means smooth functions on the dual space to
$\widetilde{\mathfrak{g}}$
serves as a phase space of integrable system.
We use the Cartan-Weyl basis in the Lie algebra and restrict the phase space to an orbit
through eliminating a subset of dynamic
variables corresponding to nilpotent commuting basis elements. The rest of dynamic variables
corresponding to basis elements give a parametrization of the orbit.
Another parametrization of the orbit is given by points of the spectral curve
$\det \bigl(\tens{L}(\lambda)\,{-}\,w\Ibb\bigr)\,{=}\,0$, where
$\tens{L}$ is the Lax matrix of the system.
Thus, we obtain a map between the dynamic
and the spectral variables, it is possible to make this map biunique.
The spectral variables are proven to be variables of separation.

The orbit approach allows an easy extension
to generic orbits in $\mathfrak{sl}(n)$ loop algebras. At the same time
expressions for the functions $\mathcal{A}$ and $\mathcal{B}$ giving variables of separation
acquire a reasonable meaning: they are implied by the procedure of restriction to an orbit,
and a simple mnemonic rule allows to write them immediately.


\label{lastpage}

\end{document}